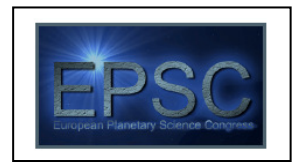

# Successful modeling of the environmental changes' influence on forests' vegetation over North Eurasia

O. Khabarova, I. Savin and M.Medvedeva

Space Research Institute (IKI) of Russian Academy of Sciences, Profsoyuznaya Str. 84/32, Moscow, 117997 Russia (olik3110@aol.com / Fax: +7-495-3331248)

## Abstract

Modeling of forests' vegetation in North Eurasia has been performed for 1982-2006 on the basis of remote sensing data. Four meteorological parameters and one parameter, characterizing geomagnetic field disturbance level, were used for this aim. It was found out that revealed formula is adequate both for coniferous evergreen and coniferous deciduous forests for accuracy to a coefficient. The most proper parameters' combination gives the correlation coefficients ~ 0.9 between modeling parameter and original data rows. These results could solve problems of climate-forests feedbacks' investigations and be useful for dendrological aims.

## 1. Introduction

Last decades are characterized by significant climate changes and consequent changes in terrestrial ecosystems, in particular, forests [1]-[3]. Plants and trees are very sensitive to variations of all environmental parameters, including the geomagnetic field [4]. Exceeding of the limit of trees' adaptability can lead to tree mortality. Even moderate climate changes produce sharp variations of forests' vegetation dynamics, which can be estimated by the volume of green biomass. Since remote sensing is one of the best ways of the global control of terrestrial surface and atmospheric processes, we have used satellite data for numerical modeling of coniferous evergreen and coniferous deciduous forests' vegetation dynamics.

## 2. Data and modeling

NOAA-AVHRR satellites provide with the measurements of Normalized Difference Vegetation Index (NDVI), satisfactory representing the green biomass volume (see data and documentation on http://gimms.gsfc.nasa.gov/ndvi/ndvie/GIMMSdocumentation_NDVIe.pdf). NDVI is calculated on the basis of the spectrally-reflective surface characteristics with the spatial resolution about 8 km. NDVI was averaged for each year for 1982-2006 over all North Eurasia (GIMMS archive) and compared with the environmental parameters, most important for trees' life.

Northern Asia region is characterized by the wide variety of bioclimatic conditions and vegetation. Meanwhile, we consider here the hypothesis: if the effect of the climate changes' influence on forests really exists, it would be expressed at global scales (under large-scale time- and spacial- averaging of the key parameters).

Figure 1 and 2 show two types of green biomass growth changes, depending on time-scales. First of all, there is a long-term trend, which demonstrates fall of forests green biomass for 25 years (possibly related to global warming). Then, relatively short variations of forests productivity with two-five years periods are observed.

A method of composite function has been used for the modeling. It was found out that relation of NDVI annual maximum to vegetation season duration, characterizing biomass variability *max-ndvi/veg.seas.duration* (speed of biomass changes, averaged for vegetation season), can be successfully revealed using five environmental parameters such as: sum of global radiation income for vegetative season

*FAR*; relative solarization for vegetation season *FAR/dur*; mean temperature for vegetation season *Temp*$_{veg}$; yearly amount of precipitation *precip*$_{all}$, and annual value of the *Kp*-index of geomagnetic activity. In the result of seeking of the various modeling functions, describing behaviour of the original time-series *max-ndvi/veg.seas.duration*, the most effective modeling parameter *PAR*$_{max/dur}$ has been found:

$$PAR_{\frac{max}{dur}} = \frac{(A \cdot Kp + B \cdot Temp_{veg}) \cdot C \cdot 10^3 + FAR/dur}{D \cdot precip_{all} 10^3 + FAR} \quad (1),$$

where *A, B, C,* and *D* are numerical coefficients (are given in Figure 1 and Figure 2).

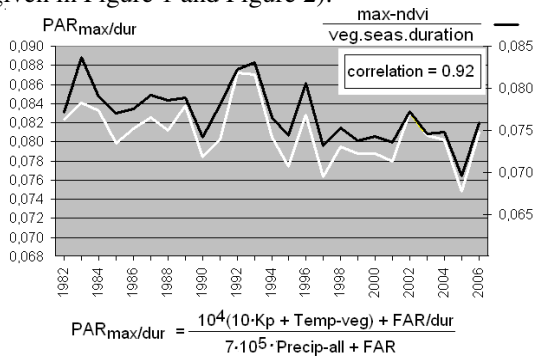

Figure 1: The result of modeling of green biomass variability (black curve - experimental data; white one - modeling parameter *PAR*$_{max/dur}$) for coniferous evergreen forests of North Eurasia for 25 years.

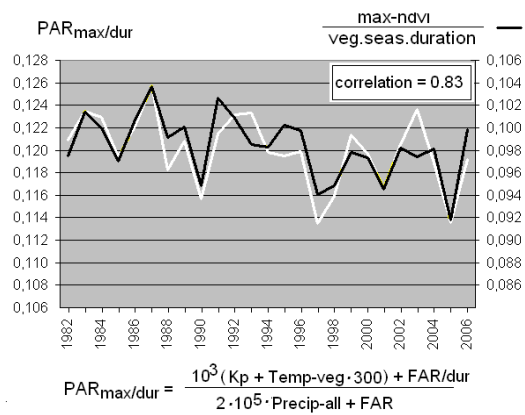

Figure 2: Like Figure 1, but for coniferous deciduous (larch) forests

Correlation coefficients between the parameter *PAR*$_{max/dur}$ and the experimental data row are correspondingly **0.92** for coniferous evergreen forests and **0.83** for coniferous deciduous (larch). It should be noted that these correlation coefficients are greater than correlation coefficients between *max-ndvi/veg.seas.duration* and parameters, separately included into the formula (1), by |0.1÷0.5|. This demonstrates success of the modeling and confirms our hypothesis about the possibility of revealing of the environmental parameters combined influence on forests.

## 3. Summary and Conclusions

The modeling results of the regional climate changes' influence on forests show that trees of North Eurasia are sensitive to variations of temperature, precipitation, solarization and geomagnetic field.
Speed of green biomass changes may be described by a function, depending on five environmental parameters both for coniferous evergreen and coniferous deciduous forests.
Coniferous deciduous forests are less sensitive to climate changes, possibly because of their natural property - very high adaptability to external conditions.
Global warming can impact the green biomass productivity negatively if the temperature increases together with the growth of precipitation.
Obtained results clearly demonstrate combined influence of environmental parameters on terrestrial ecosystem; they could explain observed episodic growth of tree-mortality rates now days and in the past, and give the possibility to predict changes in feedbacks between the Earth and forests.

## References

[1] Adams H.D., Macalady A.K., Breshears D.D., Allen C.D., Stephenson N.L., Saleska S.R., Huxman T.E., and McDowell N.G. Climate-induced tree mortality: Earth system consequences, EOS Transactions, AGU, Vol. 91, no.17, pp. 153-154, 2010.

[2] Myneni RB, Keeling CD, Tucker CJ, Asrar G, and Nemani RR., Increased plant growth in the northern high latitudes from 1981 to 1991, Nature, Vol.386 (6626), pp. 698-702, 1997.

[3] Bogaert J, Zhou L, Tucker CJ, Myneni RB, and Ceulemans R. Evidence for a persistent and extensive greening trend in Eurasia inferred from satellite vegetation index data, J. Geophys. Res., 107:10.1029/2001JD001075, 2002.

[4] Phirke P.S., Kubde A.B., and Umbarkar S.P., The influence of magnetic field on plant growth, Seed Science and Technology, V.24, N2, pp.375-392, 1996.